\documentclass[usenatbib]{basi}
\usepackage[T1]{fontenc}
\usepackage[british]{babel}
\usepackage[varg]{txfonts}
\usepackage{wrapfig}
\usepackage{lipsum}
\usepackage{xcolor}
\usepackage{rotating}
\usepackage{dcolumn}
\begin{document}
\title[BASI sample paper]{Constraining the amplitude of turbulence in solar 
corona using observations of angular broadening of radio sources}
\author[M Ingale et~al.]%
       {Madhusudan Ingale$^1$\thanks{email: \texttt{i.madhusudan@students.iiserpune.ac.in}},
       Prasad Subramanian$^{1}$ and Iver H. Cairns$^2$\\
       $^1$Indian Institute of Science Education and Research, Sai Trinity 
       Building, Pashan, Pune - 411021, India.\\
       $^2$School of Physics, University of Sydney, NSW 2006, Australia.}

\pubyear{2013}
\volume{00}
\pagerange{\pageref{firstpage}--\pageref{lastpage}}
%\status{submitted}

\date{Received --- ; accepted ---}

\maketitle
%------------------------------------------------------------------------------%
% abstract and keywords                                                        %
%------------------------------------------------------------------------------%
\label{firstpage}

\begin{abstract}
%
%We study the constraints imposed on the amplitude of density turbulence ($C_N^2$) 
%in the solar corona using observations of angular broadening of radio sources.
%We examine the connection between scattering measure and
%phase structure function to determine the role of $C_N^2$ in
%determining broadening. 
The angular broadening of compact radio sources observed through a medium having turbulent density 
irregularities is usually estimated using the phase structure function. We employ an exact formulation 
for the phase structure function that helps in obtaining an accurate estimate of angular broadening 
when the baseline lengths are comparable to the inner scale of the turbulent spectrum.
\end{abstract}


%------------------------------------------------------------------------------%
% main text of the paper, using \section, \subsection, \subsubsection          %
%------------------------------------------------------------------------------%
\section{Introduction}\label{s:intro}
Scattering of electromagnetic waves due to density turbulence in the solar corona
and the solar wind gives rise to angular broadening of the radio sources. 
%Our primary
%aim here is to use recent observations of angular broadening to investigate 
%the constraints imposed on the models of the amplitude of density turbulence $C_N^2(R)$
%in the solar corona, where R is the heliocentric distance. 
We use a formalism based on the parabolic wave approximation with general form of the phase structure function $D_{\phi}(s)$. Almost all previous investigations so far (e.g. Subramanian and Cairns 2011) have employed asymptotic forms of the phase structure function that are valid only in the limits where the baseline length is either much greater than, or much less than the inner scale of the turbulent spectrum . We use the full expression for $D_{\phi}(s)$ that does not need these limiting assumptions.\\
%We begin with the theoratical framework of phase structure function and compare the 
%predictions obtained by using two prescriptions of $C_N^2(R)$ against observations and,
%present the results of their comparison. Finally we conclude with the discussions of
%Sthese results.

\section{Phase Structure function}

An interferometer used to measure angular broadening actually measures the spatial coherence, which provides a good estimate of the mutual coherence function $\Gamma(s)$ of density turbulence through which the radiation propagates. This is directly related to the phase structure function 
$D_{\phi}(s)$ through (Ishimaru 1978, Ch 20)
\begin{equation}
\Gamma(s) = exp(-D_{\phi}(s)/2)
\label{visibility}
\end{equation}
%Alternatively, the phase structure function can be defined as (Coles and Harmon 1989)
%\begin{equation}
%D_{\phi}(s) = \langle [\phi(r) - \phi(r+s)]^2\rangle
%\label{phase}
%\end{equation}
%where $\phi$ is the phase deviation calculated along the line of sight and 
%$r$ and $s$ are transverse coordinates. 
The coherence scale $s_0$ is given by 
$D_{\phi}(s_0) = 1$. For a given wavelength $\lambda$, the extent to which an ideal point source is broadened is given by
\begin{equation} 
\theta_c = (2\pi s_0/\lambda)^{-1}
\label{eqthetac}
\end{equation}
The power in the turbulent density fluctuations in the scattering medium is assumed to follow the following spectrum (e.g., Coles et al 1987):
%In order to describe the scattering medium we use simple isotropic power spectrum with an inner scale $l_i$ as defined below, and the spatial frequency k is in radian measure (Coles et. al 1987), 
\begin{equation}
S_n(k,R) = C_N^2(R) \, k^{-\alpha} exp[-(k l_{i}/2 \pi)^{2}]
\label{spectrum}
\end{equation} 
The power spectrum thus comprises a power law with index $\alpha$ together with an exponential turnover at the inner scale $l_{i}$.\\
The general form of the phase structure function for the power spectrum (\ref{spectrum}) (Coles et al 1987), 
including the effects of a spatially varying plasma frequency $f_p(R)$ (Cairns, 1998) is given by,
\begin{equation}
D_{\phi}(s) = \frac{8\pi^2r_e^2\lambda^2}{2^{\alpha-2}(\alpha-2)}
\Gamma\left(1-\frac{\alpha-2}{2}\right)\int_{R_0}^{R_1}\! \frac{C_N^2(R)}{(1-f_p^2(R)/f^2)}
       l_i(R)^{\alpha-2}\\
   % split here
 \times\left\lbrace_1F_1\left[-\frac{\alpha-2}{2},1,-\left(\frac{s}{l_i(R)}\right)^2
 \right]-1  \right\rbrace \, \mathrm{d}R
\label{eqcsf}
\end{equation}

Here $_1F_1$ denotes the confluent hypergeometric function, $f_p$ is the plasma frequency, $f$ is the radiation frequency corrosponding to the wavelength $\lambda (= 2\pi c/ f)$, $r_e$ is the classical electron radius. 
The following limiting forms of equation~(\ref{eqcsf}) are commonly used (e.g. Coles et. al 1987, Subramanian \& Cairns, 2011):
\begin{equation}
D_{\phi}(s) = \frac{8 \pi^2 r_e^2 \lambda^2}{2^{\alpha-2}(\alpha-2)}
\frac{\Gamma\left(1-(\alpha-2)/2\right)}{\Gamma\left(1+(\alpha-2)/2\right)}
s^{\alpha-2}\int_{R_0}^{R_1}\! \frac{C_N^2(R)}{(1-f_p^2(R)/f^2)} \, \mathrm{d}R \, , \,\,\,\,s \gg l_{i}
\label{eqsgli}
\end{equation}
\begin{equation}
D_{\phi}(s) = \frac{4\pi^2 r_e^2 \lambda^2}{2^{\alpha-2}}
\Gamma\left(1-\frac{\alpha-2}{2}\right)s^2
\int_{R_0}^{R_1}\! \frac{C_N^2(R)}{(1-f_p^2(R)/f^2)}l_i(R)^{\alpha-4} \, \mathrm{d}R \, , \,\,\,\,s \ll l_{i}
\label{eqslli}
\end{equation} 
The integration limits ranges from the source ($R_0$) to the observer ($R_1$) when considering spaherical wave propagation. For plane wave propagation the integral along the line of sight can be replaced simply with the integrand multiplied by $\Delta L$, the thickness of the scattering screen.
In what follows, we compute scatter broadening angles $\theta_c$ using the full expression (eq~\ref{eqcsf}) and compare them with those obtained with the limiting expressions of eqs~(\ref{eqsgli}) and (\ref{eqslli}).

%\subsection{Amplitude of Density turbulence : $C_N^2$}

%------------------------------------------
%\begin{wrapfigure}[15]{r}{7.0cm}
%\includegraphics[width=7cm]{ampltd.eps}
%\caption{$C_N^2(R)$ as a function of heliocentric distance. 
%The dash-dotted curve is for model (A) $C_N^2(R)$ model and the The solid curves 
%is for model (B)
%}
%\end{wrapfigure}
%------------------------------------------
We use the following model for the amplitude of density turbulence, $C_N^2(R)$, first proposed by Spangler and Sakurai (1995) and later improved on by Spangler et al (2002), which was obtained from a linear fit to VLBI data of the scattering measure between $10 R_{\odot} - 50 R_{\odot}$
\begin{equation}
C_N^2 = 1.8 \times 10^{10}\left(\frac{R}{10R_{\odot}}\right)^{-3.66} 
\label{eqcnsq}
\end{equation}
The dimensions of $C_N^2(R)$ depend on $\alpha$, being m$^{-\alpha-3}$.

%The inner scale is modeled using the Coles and harmon's 1989 model, which is presumed to 
%arise from cyclotron damping (Coles and Harmon, 1989)

\section{Results and Discussion}

\begin{figure}
\centerline{\includegraphics[width=7cm]{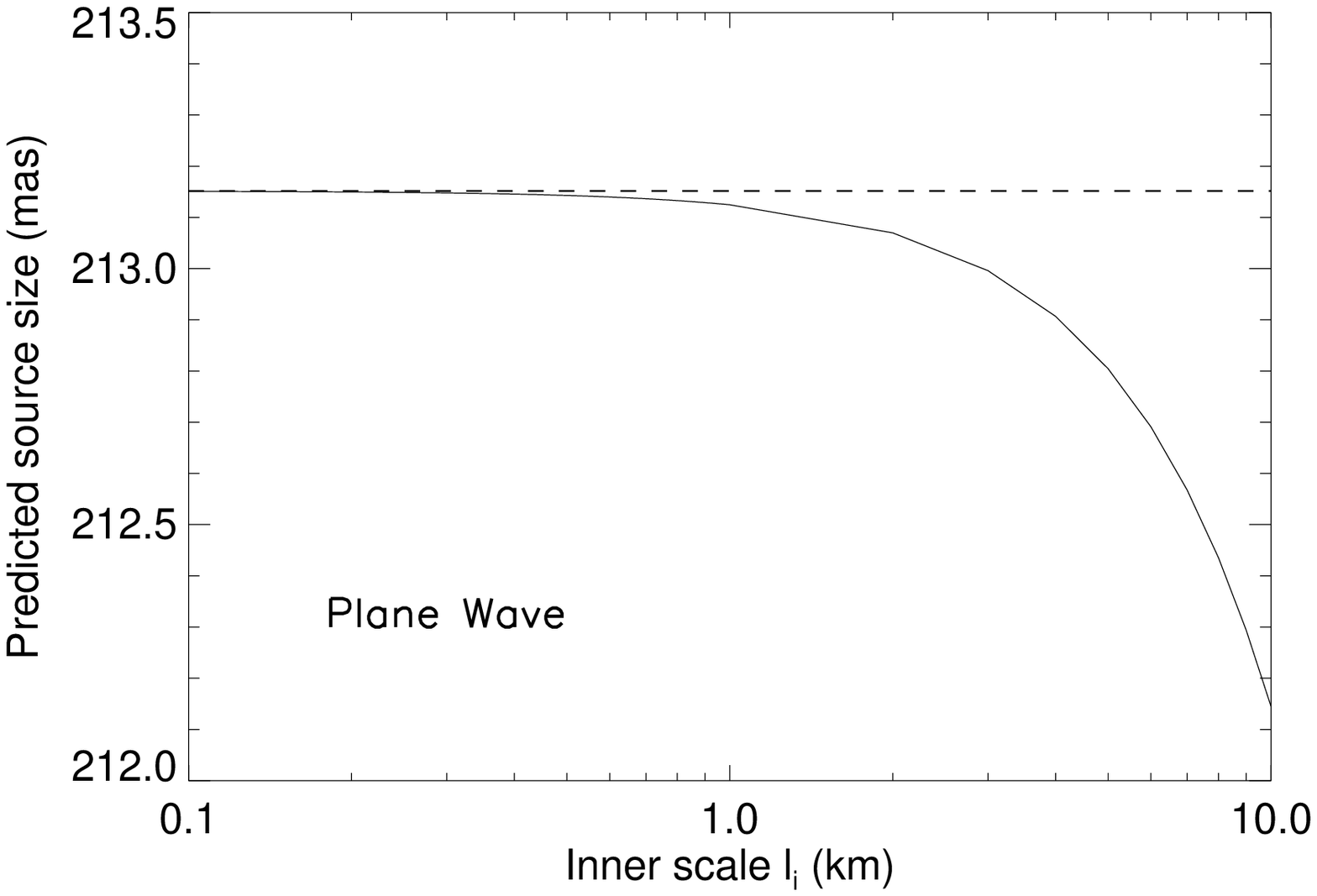}  \qquad
            \includegraphics[width=7cm]{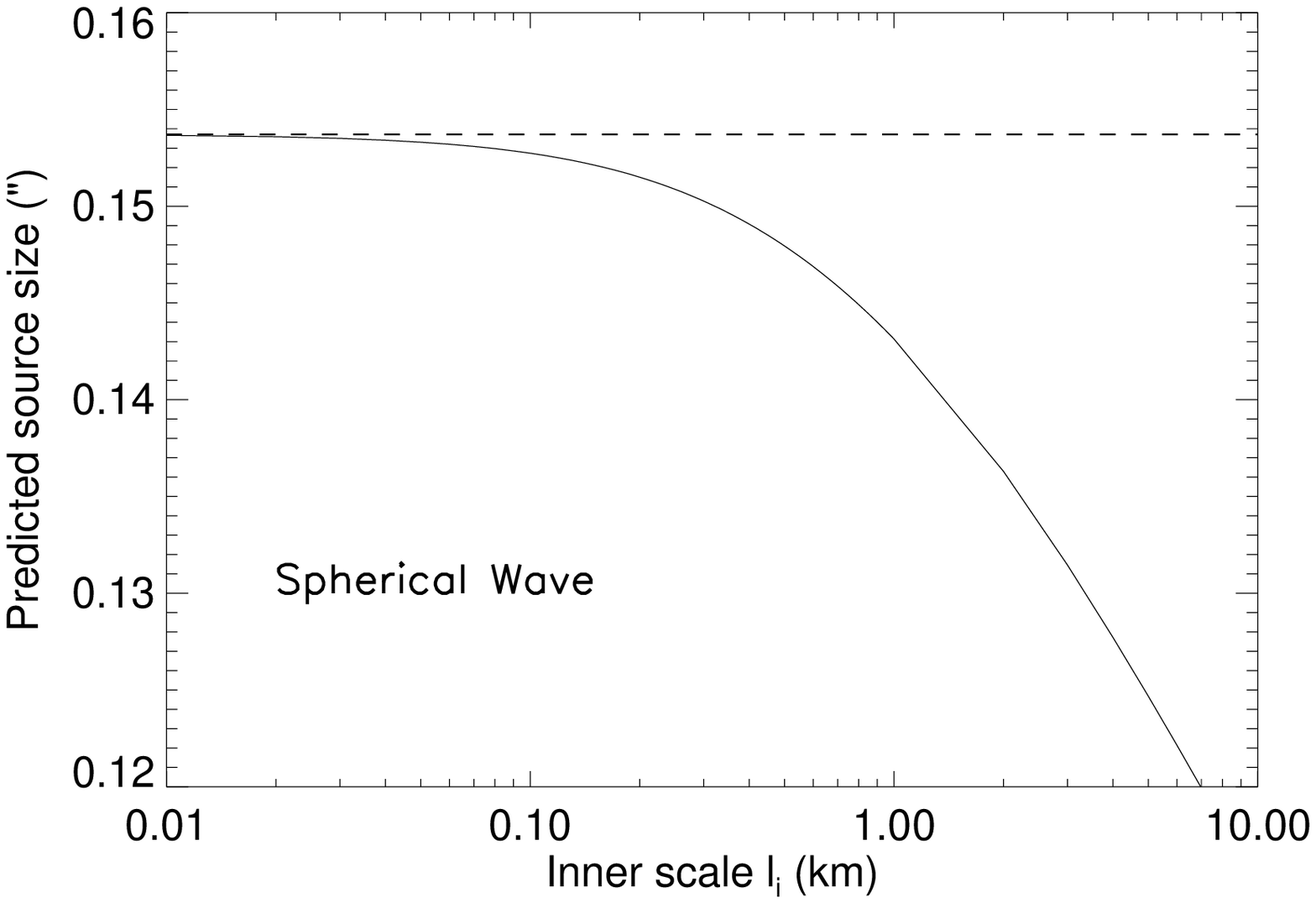}}
\caption{Predicted $\theta_c$ as a function of $l_i$ (in km). The solid line is for the full phase structure function (equation~\ref{eqcsf}) and the dashed line is for the asymptotic branch $s \gg l_i$
(equation~\ref{eqsgli}).}
\medskip
\centerline{\includegraphics[width=7cm]{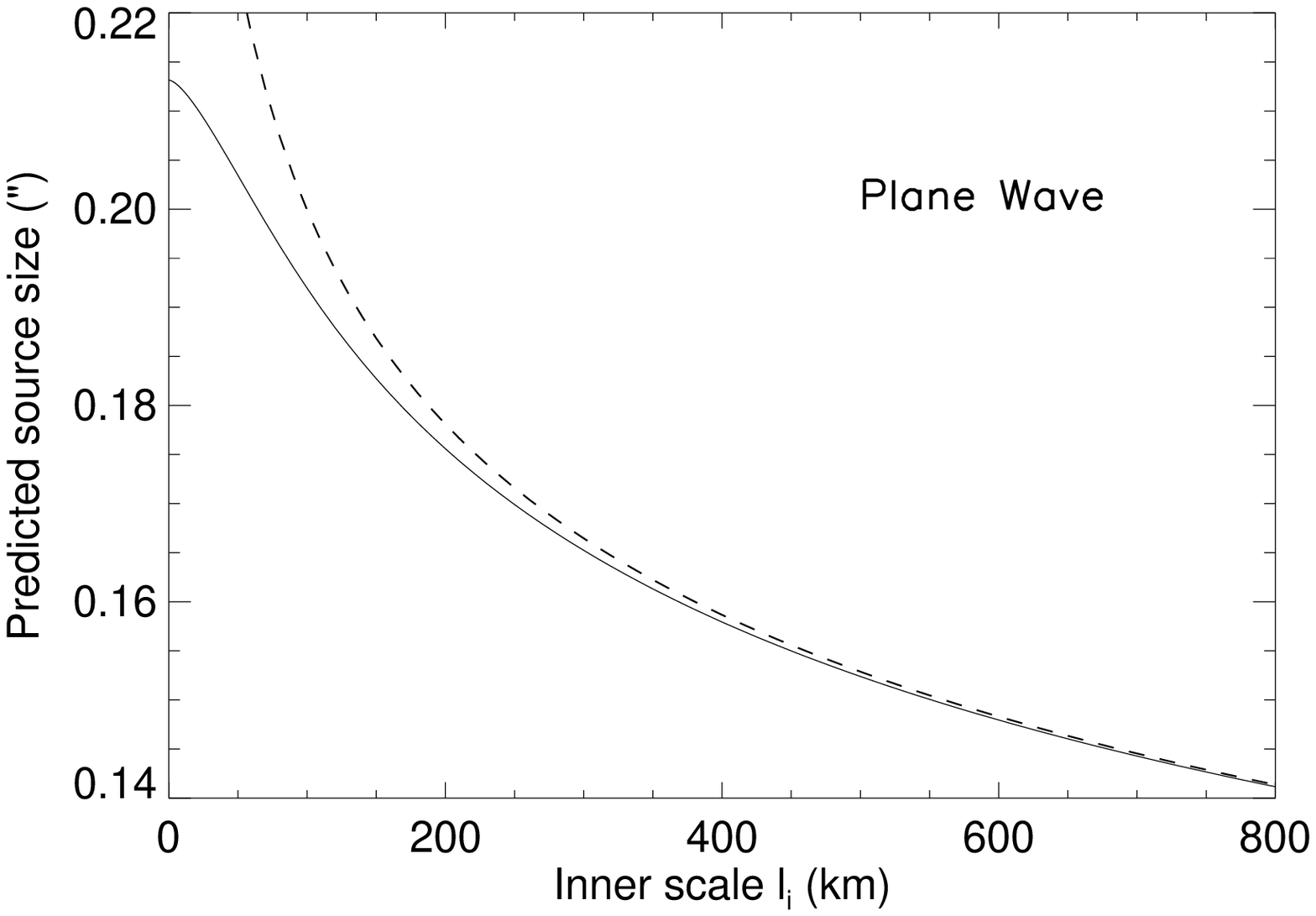} \qquad
            \includegraphics[width=7cm]{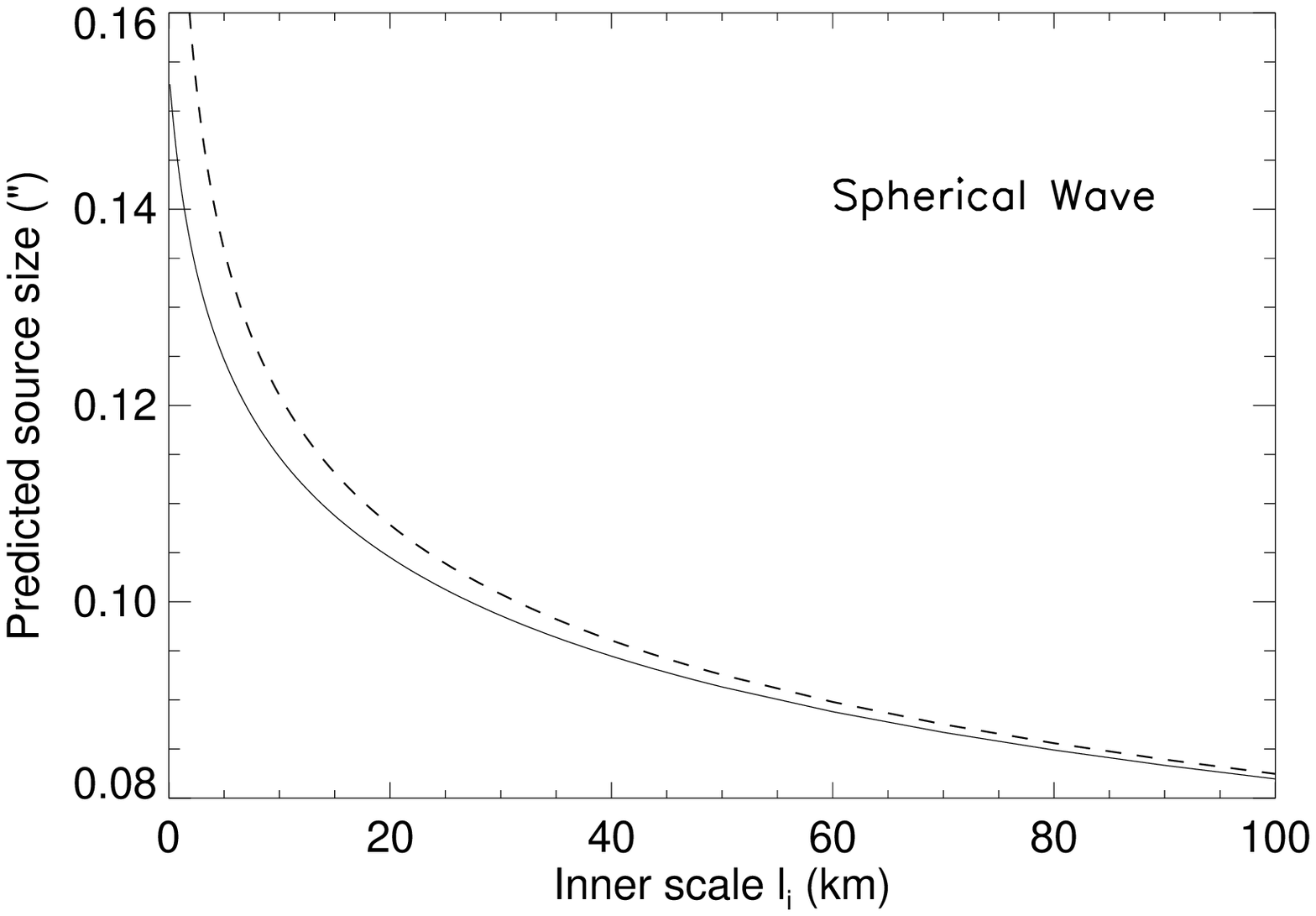}}
\caption{Predicted $\theta_c$ as a function of $l_i$ (in km). The solid line is for the general phase structure function (equation~\ref{eqcsf}) and the dashed line is for the asymptotic branch $s \ll l_i$ (equation~\ref{eqslli}).}
\end{figure}

%\subsection{Phase Structure Function}
We have treated plane wave propagation (which is appropriate for distant sources) as well as spherical wave effects (which is appropriate for sources embedded in the solar corona). 
We adopt the fourfold Newkirk model for ambient electron density (Newkirk, 1961). The fundamental emission at 327 MHz emanates from the heliocentric distance of $1.108 R_{\odot}$ with this model. For spherical wave propagation the lower limit of integration $R_0$  is taken to be $1.109 R_{\odot}$ to avoid the singularity. For the plane wave case we evaluated equation~(\ref{eqcsf}, \ref{eqsgli} and \ref{eqslli}) at an elongation of 10$R_{\odot}$, using $\Delta L = 0.5\,R_{\odot}$ for the thickness of the scattering screen. 
%We have also treated refractive index effects (Cairns 1998) and fix $\alpha$ at $11/3$ (which corresponds %to the Kolmogorov scaling).
%With this model the fundamental emission at 327 MHz emanates at $0.108 R_{\odot}$ above photosphere. 
The scattering angle $\theta_c$ is calculated using equation~(\ref{eqthetac}) at wavelength $\lambda = 91$cm, corrosponding to the fundamental emission at 327MHz. When using the full form of the phase structure function (equation~\ref{eqcsf}), we numerically determine the value of $s_{0}$ which satisfies $D_{\phi}(s_{0}) = 1$. 

We compare the prediction of the scattering angle obtained by using the full phase structure function (equation~\ref{eqcsf}) with those predicted by the asymptotic branches (equations~\ref{eqsgli} and~\ref{eqslli}). We emphasize that the quantity $l_{i}$ is varied as a free parameter in order to make this comparison. We find that the range 0.1 cm $ < l_{i} < $ 800 km spans the limits $s \ll l_{i}$ to $ s \gg l_{i}$. 

%To observe the behavior of the general phase structure function in the limits $s \ll l_{i}$ and $s \gg l_{i}$, we vary $l_i$ from 0.1 cm to 800 km.
%figures were here
Our results are summarized in Figures 1 and 2. The left panel of figure 1 shows that for plane wave propagation the predictions of equation~(\ref{eqcsf}) and the asymptotic branch for $s \gg l_{i}$ (equation~\ref{eqsgli}) agree for $l_{i} \leq 300$ m. For values of $l_{i} > 300$ m, the approximation $s \gg l_i$ is clearly not adequate, and the full structure function (equation~\ref{eqcsf}) is the appropriate expression to use.
On the other hand, the left panel of figure 2 shows that the prediction of the full structure function (equation~\ref{eqcsf}) agrees with that of the $s \ll l_{i}$ asymptotic branch (equation~\ref{eqslli}) for $l_i \geq 200$ km.

For spherical wave propagation we need to use the effective baseline $s_{eff} = sR/R_0$,
where $R_0$ is the distance of the scattering screen from the source. The right panel of Figure 1 shows that the full structure function (equation~\ref{eqcsf}) agrees with that of the $s_{eff} \gg l_{i}$ branch (equation~\ref{eqsgli}) for $l_{i} < 100$ m, while the right panel of Figure 2 shows that the full structure function prediction agrees with the $s_{eff} \ll l_{i}$ branch (equation~\ref{eqslli}) for $l_i > 60$ km.

%Coles and Harmon (1989) found that for the proton number density $N = 1.3\times 10^{4}$ cm$^{-3}$ inner scale at $8 R_{\odot}$ is 6 km which roughly agrees with the observations. We find the region of inner scale where the full phase structure function is the most relevant is between 100 m $< l_i < 40$ km, i.e. bellow $50 R_{\odot}$ the obsrvational values of the inner scale lies in exactly the region for which asymptotic branches are inadequate and the full phase structure function should be used.

To summarize, we note that the predictions of the asymptotic branches fail for $300$ m $< l_i < 200$ km for plane wave propagation. For spherical wave propagation, this range is $100$ m $< l_i < 60$ km. The full structure function needs to be used under these circumstances.

This work assumes that the spatial power spectrum of density turbulence is isotropic, whereas it is well known that the power spectrum is anisotropic close to the Sun (e.g., Narayan et al 1989; Armstrong et al 1990). Further work is needed to incorporate anisotropy (e.g., Backer \& Chandran 2002). Although we have so far taken the power law index $\alpha$ in the turbulent density spectrum (Eq~\ref{spectrum}) to be equal to the Kolmogorov value of 11/3, it may be noted that there is evidence for a flattening of the spectrum to power law indices closer to 3 between scales ranging from around 100 km to the inner scale at heliocentric distances of a few $R_{\odot}$ (Coles \& Harmon 1989). The extent of this flattening strongly depends upon the phase of the solar cycle and the speed of the solar wind in question (Manoharan 1994), but it is worth examining this issue by using $\alpha = 3$ in Eq~(\ref{spectrum}), as in Bastian (1994).  When using $\alpha = 3$, the normalization for $C_{N}^{2}$ in Eq~(\ref{eqcnsq}) changes from 1.8 $\times 10^{10}$ to $10^{12}$. With $\alpha = 3$, for plane wave propagation (at an elongation of 10 $R_{\odot}$ and using a screen thickness $\Delta L = 0.5\,R_{\odot}$ as before) the predictions of the combined structure function and the asymptotic branches disagree for values of the inner scale in the range $100\, m < l_{i} < 1000\, km$. Using $\alpha = 3$ for spherical wave propagation, the disagreements between the combined structure function predictions for angular broadening observed at the Earth and those of the asymptotic branches are observed for $1\, m < l_{i} < 100\, km$. 
%{\bf OR} \textit{Whereas for the spherical wave propagation the disagreement between the combined %structure function predictions for angular broadening and those of the asymptotic branches observed at the %Earth, found to be between $1\, m < l_i < 100\, km$}.

\section{Conclusion}
We have explored the broadening of an ideal point source with a specific model for the 
amplitude of turbulence $C_N^2$. 
We have examined both plane and spherical wave propagation, and find that for $100$ m $< l_i < 200$ km the predictions of the asymptotic branches are inadequate and the full phase structure function  (equation~\ref{eqcsf}) should be used in order to obtain accurate results. Coles \& Harmon (1989) note that 400 m $< l_{i} <$ 100 km for heliocentric distances ranging from 2 to 100 $R_{\odot}$; clearly, the full structure function (Eq~\ref{eqcsf}) needs to be used under these circumstances. It is also clear from Figure 1 that the value of the scatter-broadened angle is senstitive to the value of the inner scale $l_{i}$ for $l_i > 100$ m. In other words, inner scale effects are important for $l_i > 100$ m. Importantly, this is also the regime where the predictions of the $s \gg l_{i}$ branch fail, and it is essential to use the full phase structure function. We have also explored the consequences of using a power law index $\alpha = 3$ that is flatter than the Kolmogorov value for the density turbulence spectrum. We find that this change considerably extends the range of inner scales for which the asymptotic branch predictions and those using the full phase structure function disagree. This underlines the importance of using the full phase structure function for quantitative estimates of angular broadening.

%------------------------------------------------------------------------------%

\section*{Acknowledgements}

MI and PS acknowledge support from the CAWSES-India program administered by the Indian Space Research Organization. IHC acknowledges the financial support from the Australian Research Council. They thank the anonymous referee for a very helpful report that has considerably helped improve this paper. 

%------------------------------------------------------------------------------%
% bibliography: produced from ADS using custom format of                       %
%                                                                              %
%     %z132 \\bibitem[%\2%(y)%\3m]%{R}\n   %\8.1g,%\Y,%\q,%\V,%\ p             %
%------------------------------------------------------------------------------%

%\appendix
%------------------------------------------------------------------------------%
% appendices:                                                                  %
%------------------------------------------------------------------------------%
%\section{An Example Appendix}\label{a:example}

%This is an example appendix, which uses the \verb|\appendix| command, and then
%subsequently the \verb|\section{...}|, \verb|\subsection{...}| etc.\ commands.

\label{lastpage}
%------------------------------------------------------------------------------%
\end{document}